\documentclass[aps,pra,reprint,superscriptaddress]{revtex4-2}

\usepackage{graphicx}
\usepackage{dcolumn}
\usepackage{color}
\usepackage{bm}
\usepackage[bookmarksopen]{hyperref}
\usepackage{pgfplots}
\usepackage{pgfplotstable}
\usepackage{comment}
\pgfplotsset{compat=1.18}
\usepgfplotslibrary{groupplots}
\usepackage{siunitx}
\usepackage{mathtools}
\usepackage[version=4]{mhchem}

\usepackage{tikz}

\newcommand{\labeledimage}[3]{%
    \begin{tikzpicture}
        \node[anchor=south west, inner sep=0] (img) at (0,0)
            {\includegraphics[width=#1]{#3}};
        \node[anchor=north west, xshift=-5pt, yshift=0pt, fill=white,
              fill opacity=0.7, text opacity=1, inner sep=2pt]
            at (img.north west) {#2};
    \end{tikzpicture}%
}
\newcommand{\labeledimagetwo}[6]{%
    \begin{tikzpicture}
        \node[anchor=south west, inner sep=0] (img) at (0,0)
            {\includegraphics[width=#1]{#2}};
        \node[anchor=west, xshift=-5pt, fill=white,
              fill opacity=0.7, text opacity=1, inner sep=2pt]
            at ($(img.north west)!#4!(img.south west)$) {#3};
        \node[anchor=west, xshift=-5pt, fill=white,
              fill opacity=0.7, text opacity=1, inner sep=2pt]
            at ($(img.north west)!#6!(img.south west)$) {#5};
    \end{tikzpicture}%
}

\definecolor{darkblue}{rgb}{0.0,0.0,0.4}
\definecolor{darkgreen}{rgb}{0.0,0.4,0.0}
\definecolor{darkred}{rgb}{0.6,0.0,0.0}
\hypersetup{colorlinks,linkcolor=darkblue,citecolor=darkblue,urlcolor=darkblue}

\begin{document}
\title{Kinetic modeling of molecular beam formation in a cryogenic buffer-gas cell}
\author{Ekkehard Steinmacher}
\affiliation{Vienna Center for Quantum Science and Technology, Atominstitut, TU Wien,  Stadionallee 2,  1020 Vienna,  Austria}
\affiliation{LSS, Faculty of Engineering, Friedrich-Alexander-Universität Erlangen-Nürnberg, Germany}
\author{Phillip Groß}
\affiliation{Vienna Center for Quantum Science and Technology, Atominstitut, TU Wien,  Stadionallee 2,  1020 Vienna,  Austria}

\author{Franziska Tuttas}
\affiliation{Institute of Space Systems, University of Stuttgart, Germany}

\author{Marcel Pfeiffer}
\affiliation{Institute of Space Systems, University of Stuttgart, Germany}

\author{Tim Langen}
\affiliation{Vienna Center for Quantum Science and Technology, Atominstitut, TU Wien,  Stadionallee 2,  1020 Vienna,  Austria}
\email{tim.langen@tuwien.ac.at}

\begin{abstract}
Cryogenic buffer-gas cells are widely used to produce cold molecular beams, but the microscopic dynamics governing beam formation remain challenging to model. Here we present fully kinetic simulations of a cryogenic buffer-gas cell using the Direct Simulation Monte Carlo method implemented in the PICLas framework, treating the buffer gas and ablated molecules within a single unified model. We capture characteristic features of cryogenic buffer-gas sources, including plume cooling, directed transport toward the aperture, and the formation of a slow molecular beam, while also resolving energy transfer from the hot ablation plume to the helium buffer gas that is inaccessible to existing approaches relying on the background-gas approximation. Our results demonstrate that fully kinetic simulations can provide detailed insights into buffer-gas cell dynamics and open a route toward a systematic optimization of such sources.
\end{abstract}

\maketitle

\section{Introduction}

Cold molecular beams produced in cryogenic buffer-gas cells have become an important tool in atomic and molecular physics~\cite{Hutzler2012, Hemmerling2014, truppeBufferGasBeam2018, Barry2011}. By introducing hot atoms or molecules into a cryogenic environment filled with an inert buffer gas, the species of interest rapidly thermalize through collisions and can be extracted through an aperture to form a cold, slow beam. Such sources have enabled a wide range of experiments, including precision spectroscopy, studies of cold and ultracold collisions, and laser cooling of molecules~\cite{Spaun2016, ACME2018, Miyamoto2022, Fitch2021,Langen2023}.

Several numerical studies have modeled the dynamics of molecules in buffer-gas cells, using continuum fluid simulations, random-walk trajectory methods, and combinations of
these~\cite{takahashiSimulationCryogenicBuffer2021, gantnerBuffergasCoolingMolecules2020, vogeleyHydrodynamicEffectsCryogenic2025}. Kinetic Monte Carlo methods have also been applied to trace molecules with state-resolved collision cross sections through a uniform helium bath~\cite{doppelbauerUsingDirectSimulation2017}. A common feature of all of these approaches is that the buffer-gas flow and dynamics of the molecular trace species are treated separately. In this background-gas approximation, the buffer-gas flow field is either pre-computed or assumed to be unperturbed, and the molecular trajectories are subsequently propagated through it. While this separation has proven useful for exploring source designs~\cite{takahashiSimulationCryogenicBuffer2021, vogeleyHydrodynamicEffectsCryogenic2025}, it excludes a self-consistent description of the coupled dynamics between the molecules and the buffer gas. In particular, the assumption that the buffer gas remains undisturbed by a hot molecular plume rules out any description of energy transfer from the molecules to the helium, an effect that may be relevant under typical experimental conditions, in which the molecules are created in situ using laser ablation~\cite{skoffDiffusionThermalizationOptical2011}.

Here we perform fully kinetic simulations of a cryogenic buffer-gas cell using the Direct Simulation Monte Carlo method, implemented in the PICLas framework. Within this framework, the helium buffer gas and the ablated molecules can be treated within a single unified model without invoking a separation between flow and particle dynamics. Unlike existing approaches, this unified treatment naturally captures the mutual interaction between the buffer gas and molecular species, allowing both established beam-formation processes and coupled dynamical effects to be investigated. In doing so, our simulations allow us to capture the helium flow, the expansion, and thermalization of a molecular plume following its creation by ablation, and the formation of the molecular beam within a single consistent framework.

\begin{figure}[t]
    \centering
    \includegraphics[width=0.46\textwidth]{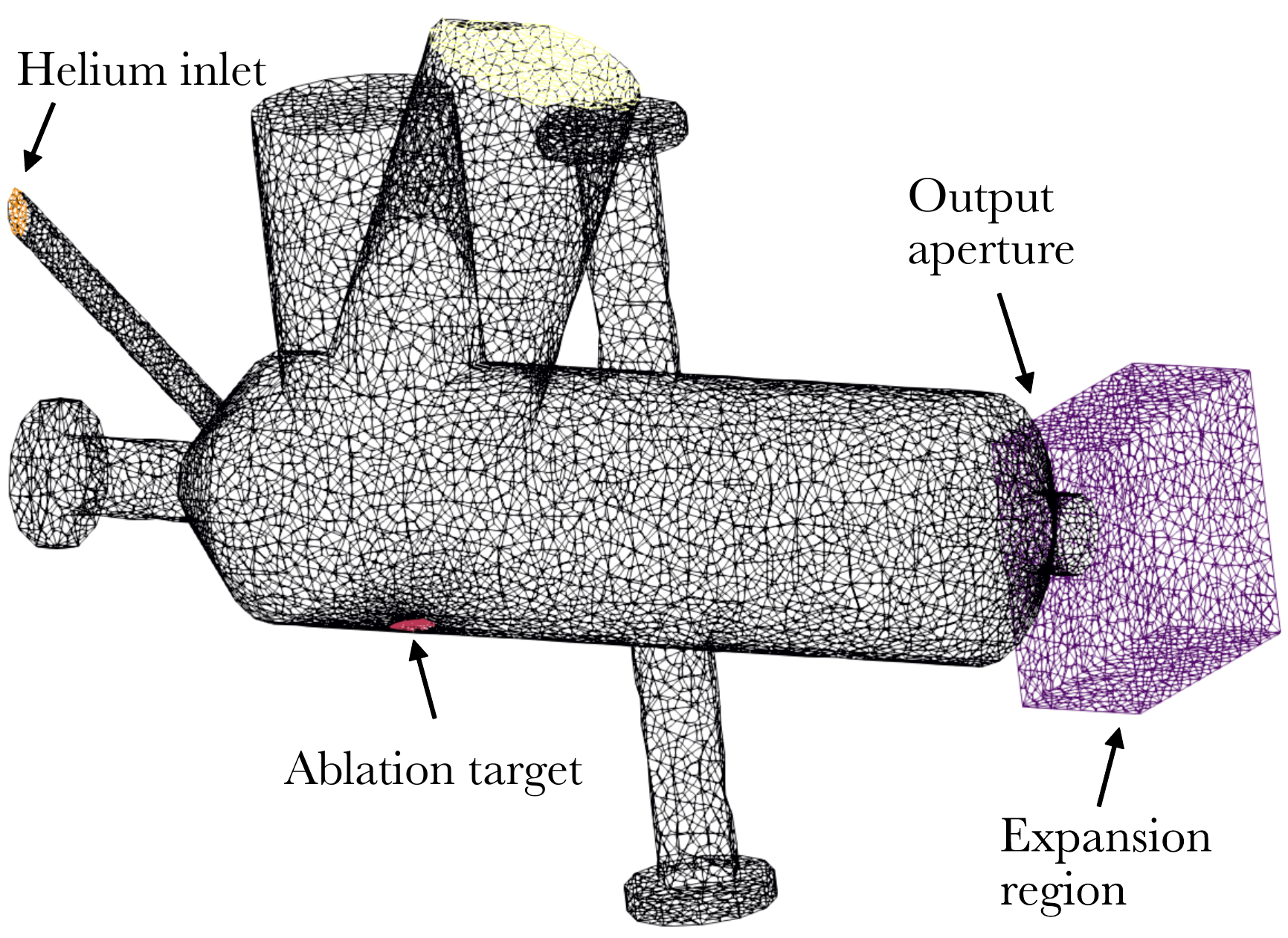}
    \caption{Simulated buffer-gas cell geometry. A helium buffer gas enters through an inlet, molecules are created by ablation from a target, and both buffer gas and molecules expand through an aperture into an expansion region, where the molecular beam formation can be analyzed. The different colors indicate the various boundary types, including the helium inlet (orange), the helium-reflecting and molecule-absorbing cell walls (black), and the open boundaries through which particles
    leave the computational domain (purple). In the simulation, the cell interior and the downstream expansion region are resolved by an unstructured hexahedral mesh. The use of such a mesh permits the simulation of complex buffer-gas cell geometries without modifications to the underlying kinetic model, making the framework readily adaptable to other source designs.}
    \label{fig:bgc_hex_mesh_visu}
\end{figure}

To connect simulation results directly to experimental observables, we use tracing planes, a new diagnostic feature in \mbox{PICLas} that records the full particle state at any specified position in the simulation domain. We demonstrate this approach in a representative buffer-gas cell geometry using helium (He) as the buffer-gas species and calcium monofluoride (CaF) as the example molecular species. The latter is widely used in molecular laser cooling experiments~\cite{Truppe2017, Anderegg2019, hollandOndemandEntanglementMolecules2023} and is therefore of particular interest. However, the mechanisms and conclusions generalize to a wide range of other molecular species.


\section{Simulation framework}
\label{sec:framework}

In a cryogenic buffer-gas cell, the flow conditions span a wide range of regimes within a single device. The buffer-gas density varies by orders of magnitude between the cell interior and the expanding beam downstream from the cell aperture. As a result, the corresponding Knudsen numbers $\mathrm{Kn} = \lambda / d_\mathrm{aperture}$, with the local mean free path $\lambda$ and the cell aperture diameter $d_\mathrm{aperture}$, range from approximately $0.01$ inside the cell at high flows to values well above unity in the expansion region.
Similarly, the Reynolds numbers $\mathrm{Re} = \rho w d_\mathrm{aperture} / \eta$, with the local density $\rho$, the flow velocity $w$ and the dynamic viscosity $\eta$, range from below unity in the effusive regime to $\sim 100$ near the supersonic regime for typical buffer-gas flow rates. No single continuum description is valid across this range and, therefore, a kinetic treatment based on the Boltzmann equation is required.

\subsection{PICLas framework}
We solve the Boltzmann equation numerically using the Direct Simulation Monte Carlo (DSMC) method \cite{birdMolecularGasDynamics1994}. In DSMC, the phase-space distribution is represented by an ensemble of simulation particles, each statistically representing many real particles. The ensemble reproduces the underlying distribution functions of the real particles, while the reduced number of particles makes the simulations computationally tractable. Within each time step, particle motion and collisions are treated as decoupled processes. Particles first propagate ballistically according to their current velocities, boundary conditions are applied where trajectories intersect cell walls, and collision partners are then selected stochastically within spatial cells. The collision step updates both translational velocities and internal energy states. In all simulations presented, the collision-cell size remains below the local mean free path, and the time step below the local mean collision time throughout the computational domain, satisfying the standard DSMC validity criteria~\cite{Binder2018}.

Macroscopic quantities such as density, temperature, and flow velocity are recovered by averaging over the particle ensemble. A key advantage of this approach is that it remains valid across flow regime transitions within a single simulation --- the same framework applies in the dense, nearly hydrodynamic interior of the cell, at the aperture where strong density gradients develop, and in the collisionless expansion beyond the aperture.

Our simulations are performed using PICLas~\cite{ortweinPiclasHighlyFlexible2017, pfeifferDirectSimulationMonte2016, nizenkovModelingChemicalReactions2017,Fasoulas2019}, a high-performance framework originally developed for aerospace applications involving rarefied and plasma flows, designed for use in high-performance computing clusters. The framework is highly flexible and adaptable to different molecular species, buffer gases, and cell geometries, as discussed in the following. To the best of our knowledge, our work represents the first application of PICLas to atomic and molecular physics. 

A typical pure helium flow simulation uses roughly $10^7$ simulation particles and requires $10^2$ to $10^4$ CPU hours, depending on the flow rate. Simulations of the molecular motion in the background-gas approximation are comparably inexpensive (on the order of $10^2$ CPU hours), allowing different molecular initial configurations to be explored efficiently, while fully coupled simulations are as expensive as pure helium simulations.

\begin{figure}[t]
    \centering
    \includegraphics[width=0.45\textwidth]{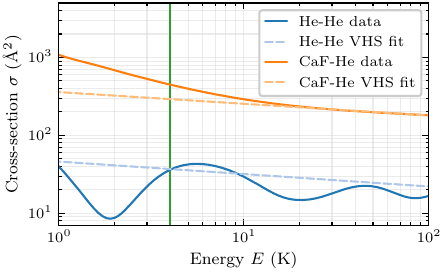}
    \caption{
    Elastic collision cross-sections for He--He (blue) and CaF--He (orange) interactions, together with VHS model fits (dashed lines)~\cite{londonoRotationalQuenchingMonofluorides2025}. The He--He cross-section is accurately reproduced by standard PICLas parameters above \SI{4}{K} (vertical green line). For CaF--He, the fit is weighted toward the high-energy regime, which governs the early thermalization of the hot ablation plume and is most relevant for the dynamics studied here. More complex collision models could be employed within the same framework.
}
    \label{fig:CaFHe_HeHe_XS_VHS_Model}
\end{figure}

\subsection{Particle interactions}
In the simulations, elastic particle interactions are described using the variable hard sphere (VHS) model~\cite{birdMolecularGasDynamics1994}, in which the collisional cross-section depends on the relative velocity of the colliding pair as
\begin{equation}
    \sigma_\mathrm{VHS} = \pi d_\mathrm{ref}^2
    \left[ \frac{2(2-\omega_\text{P})\,k_\mathrm{B} T_\mathrm{ref}}{\mu v^2} \right]^{\omega_\text{P}},
    \label{eq:vhs}
\end{equation}
where $d_\mathrm{ref}$ is an effective diameter, $\mu$ the reduced mass of the colliding pair, $v$ their relative velocity, $T_\mathrm{ref}$ a reference temperature, and $\omega_\text{P} = \omega - 0.5$, with $\omega$ being the viscosity index. For He--He interactions, we use standard tabulated parameters ($T_\mathrm{ref} = 273\,\mathrm{K}$, $d_\mathrm{ref} = 2.33 \times 10^{-10}\,\mathrm{m}$, $\omega = 0.66$) \cite{birdMolecularGasDynamics1994}, which accurately reproduce the ab-initio He--He elastic cross sections above 4\,K. The collisional properties of our representative molecular species CaF with He have been calculated ab initio~\cite{londonoRotationalQuenchingMonofluorides2025}. For CaF--He interactions, VHS parameters are fitted to these ab-initio cross sections, with the fit weighted toward the high-energy tail that governs the early thermalization of the ablation plume. The CaF--He parameters are $T_\mathrm{ref} = 4\,\mathrm{K}$, $d_\mathrm{ref} = 9.21 \times 10^{-10}\,\mathrm{m}$, and $\omega = 0.65$. The resulting cross sections and their VHS fits are shown in Fig.~\ref{fig:CaFHe_HeHe_XS_VHS_Model}. Because the thermalization process spans a wide range of collision energies, the adopted fit represents a compromise between accurately describing the early high-energy collisions and the later stages of cooling.

Importantly, the VHS model only describes elastic collisions. For molecular species, inelastic collisions can additionally transfer energy into or out of the internal vibrational and rotational degrees of freedom. In our simulations, the corresponding relaxation processes are treated using fixed  empirical per-collision relaxation probabilities, as discussed further in Sec.~\ref{sec:molecules}.

More accurate interaction models could, in principle, be employed within the same framework at the cost of increased computational effort, for example using state-resolved inelastic cross sections~\cite{doppelbauerUsingDirectSimulation2017} or potential-energy-surface-based scattering models. The computational demands of such models can be readily accommodated on our high-performance computing infrastructure, making their implementation feasible within the present framework. However, the interaction model adopted here is already sufficient to demonstrate the basic principles and capabilities of our framework. More sophisticated interaction models and quantitative comparisons with dedicated experiments will be the subject of future work.

\subsection{Cell geometry and boundary conditions}

The geometry of the buffer-gas cell is represented by an unstructured hexahedral mesh. The simulated cell is based on our typical experimental design~\cite{Albrecht2020,Rockenhaeuser2023} and features a complex geometry that includes several side chambers that are used to introduce probe laser beams as well as the buffer gas. The use of an unstructured mesh allows such complex geometries to be represented without modifications to the underlying DSMC algorithm, making the framework readily adaptable to alternative cell and aperture designs.  

The simulation domain includes the full cell interior and a rectangular expansion volume downstream of the aperture, allowing the molecular beam to be followed into the free-expansion region (see Fig.~\ref{fig:bgc_hex_mesh_visu}). 

Helium is injected at the buffer-gas inlet with a specified surface flux corresponding to the desired flow rate. Cell walls are treated as diffuse reflectors for the buffer gas and as fully absorbing boundaries for molecules, consistent with the negligible vapor pressure of the molecular species at cryogenic temperatures. The exterior boundaries of the expansion volume are open, allowing particles to leave the simulation domain freely. This volume is also chosen sufficiently large for the gas to approach the fully ballistic, non-collisional regime at the boundary. Molecules are introduced at the target through a surface flux as a thermal ensemble, following established models of laser ablation~\cite{taralloBaHMolecularSpectroscopy2016}.

\subsection{Tracing planes and observables}
To connect the simulation output to typical experimental observables, we use tracing planes as a particle-tracing diagnostic tool. A tracing plane is a geometric surface --- circular or rectangular --- placed at any position within the simulation domain. Whenever a particle trajectory crosses the plane within a time step, the full particle state (position, velocity, and internal energy) is recorded. This mimics, for example, the role of a probe laser in an experiment, which interrogates the molecular ensemble at a specific location inside the cell or downstream in the beam, allowing velocity distributions,  temperatures and other typical experimental observables to be extracted at each stage of the beam formation process.

Unless stated otherwise, all reported observables are extracted from the full ensemble of recorded simulation particles that pass through such a tracing plane. Any sampling uncertainties are therefore governed by the finite number of sampled particles and scale approximately as $1/\sqrt{N}$, where $N$ is the number of contributing particles. For the molecular beam simulations presented in Sec.~\ref{sec:extraction}, beam observables are typically extracted from on the order of $10^3$ detected simulation particles, corresponding to relative statistical uncertainties of only a few percent. Systematic uncertainties associated with the collision model and the assumed initial conditions are therefore expected to exceed the remaining statistical uncertainty for most reported quantities.

\section{Helium buffer-gas dynamics}
\label{sec:helium}


\begin{figure}[t]
    \centering
    \labeledimage{0.43\textwidth}{(a)}{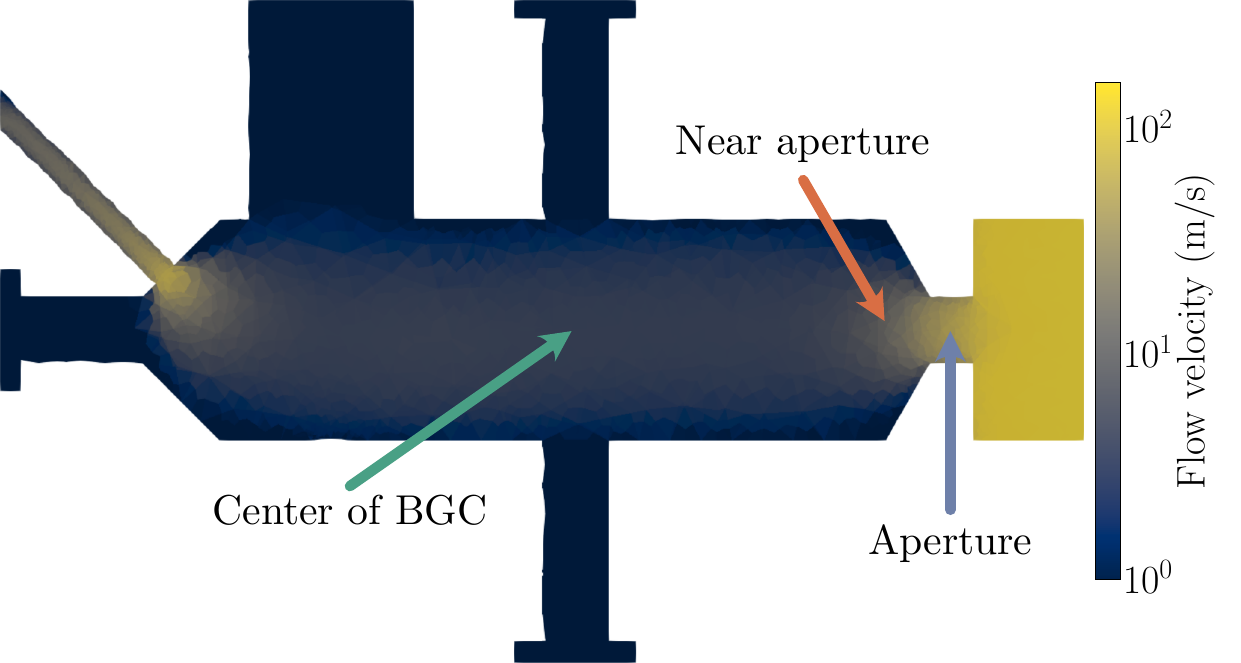} \\~\\
    \labeledimage{0.43\textwidth}{(b)}{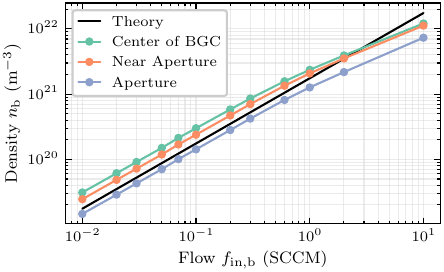}
    \caption{
   Helium flow in the buffer-gas cell in the steady state. (a) Helium flow field for an inlet flow of \SI{1}{SCCM}. The color scale indicates the magnitude of the flow velocity. The velocity remains low throughout most of the cell interior (\SI{5}{m / s}) and increases to about \SI{160}{m / s} as the gas expands through the aperture into vacuum. (b) Helium  buffer-gas density $n_\mathrm{b}$ as a function of inflow, measured at the three probe locations indicated by the arrows in panel (a). The simulated densities follow the simplified analytical prediction of Eq.~\ref{eq:steady_state_density} (black line) over more than two orders of magnitude in helium flow $f_\textrm{in,b}$. The small deviations between the simulated and analytical values reflect density gradients within the cell that are neglected in the simplified model of Eq.~\ref{eq:steady_state_density}.
      }
    \label{fig:he_flow_field}
    \label{fig:he_density_vs_flow}
\end{figure}

As a first step, we simulate the helium buffer gas alone, without any molecular species present. This allows us to verify that the simulated flow reproduces the expected behavior of a cryogenic buffer-gas cell before introducing the additional complexity of molecular thermalization and plume dynamics. 

\subsection{In-cell helium dynamics}

Helium is injected with a temperature of \SI{4}{K} and the simulation is run until the steady-state condition $f_\mathrm{out} = f_\mathrm{in}$ is reached, where the outflow through the aperture $f_\mathrm{out}$ is equal to the prescribed inlet flow $f_\mathrm{in}$. Figure \ref{fig:he_flow_field}a shows the resulting steady-state helium velocity field for a buffer-gas inflow of \SI{1}{SCCM}, where SCCM denotes standard cubic centimeters per minute or $1\,\mathrm{SCCM} \approx 4.48\times10^{17}$ particles per second.

The bulk velocity of the helium inside the main chamber is comparably low at \SI{5}{m / s}, increasing to approximately \SI{160}{m / s} after passing the output aperture. 

A commonly used estimate for the resulting in-cell helium density in such a steady state is given by~\cite{Hutzler2012}
\begin{equation}
    n_\mathrm{b} = \frac{4 f_\mathrm{in,b}}{A_\mathrm{aperture}\,\bar{v}_\mathrm{b}},
    \label{eq:steady_state_density}
\end{equation}
where $A_\mathrm{aperture} \approx 7 \times 10^{-6}\,\mathrm{m}^2$ is the aperture area for our \SI{3}{mm} diameter aperture, and $\bar{v}_\mathrm{b} = \sqrt{8 k_\mathrm{B} T_\mathrm{b} / \pi m_\mathrm{b}} \approx \SI{145}{m\per s}$ is the mean thermal helium velocity at a buffer-gas temperature $T_\mathrm{b}=4\,$K. Here, $m_\mathrm{b}$ is the mass of the helium buffer-gas atoms and $k_\mathrm{B}$ is Boltzmann's constant. Equation~\ref{eq:steady_state_density} assumes effusive flux through the aperture and a uniform in-cell density, and is therefore expected to hold best at low flows.

The simulated densities extracted at three probe positions within the cell are compared to this result in Fig.~\ref{fig:he_density_vs_flow}b and reproduce the expected linear scaling with inflow across more than two orders of magnitude between \SIrange{0.05}{10}{SCCM}, corresponding to helium densities in the range of \SIrange{e19}{e22}{m^{-3}} within the cell. 

Deviations from the analytical prediction reflect density gradients present in the cell that the uniform-density model of Eq.~\ref{eq:steady_state_density} does not capture. These are most pronounced near the aperture, where the density gradient is steepest. In addition, for inflows above approximately \SI{5}{SCCM}, the density converges more slowly toward a steady state and remains slightly below the prediction within the simulated time window, although the velocity field is already fully established~\cite{bulleidCharacterizationCryogenicBeam2013,
vogeleyHydrodynamicEffectsCryogenic2025}.

\begin{figure}[tbp]
    \centering
    \labeledimagetwo{0.45\textwidth}{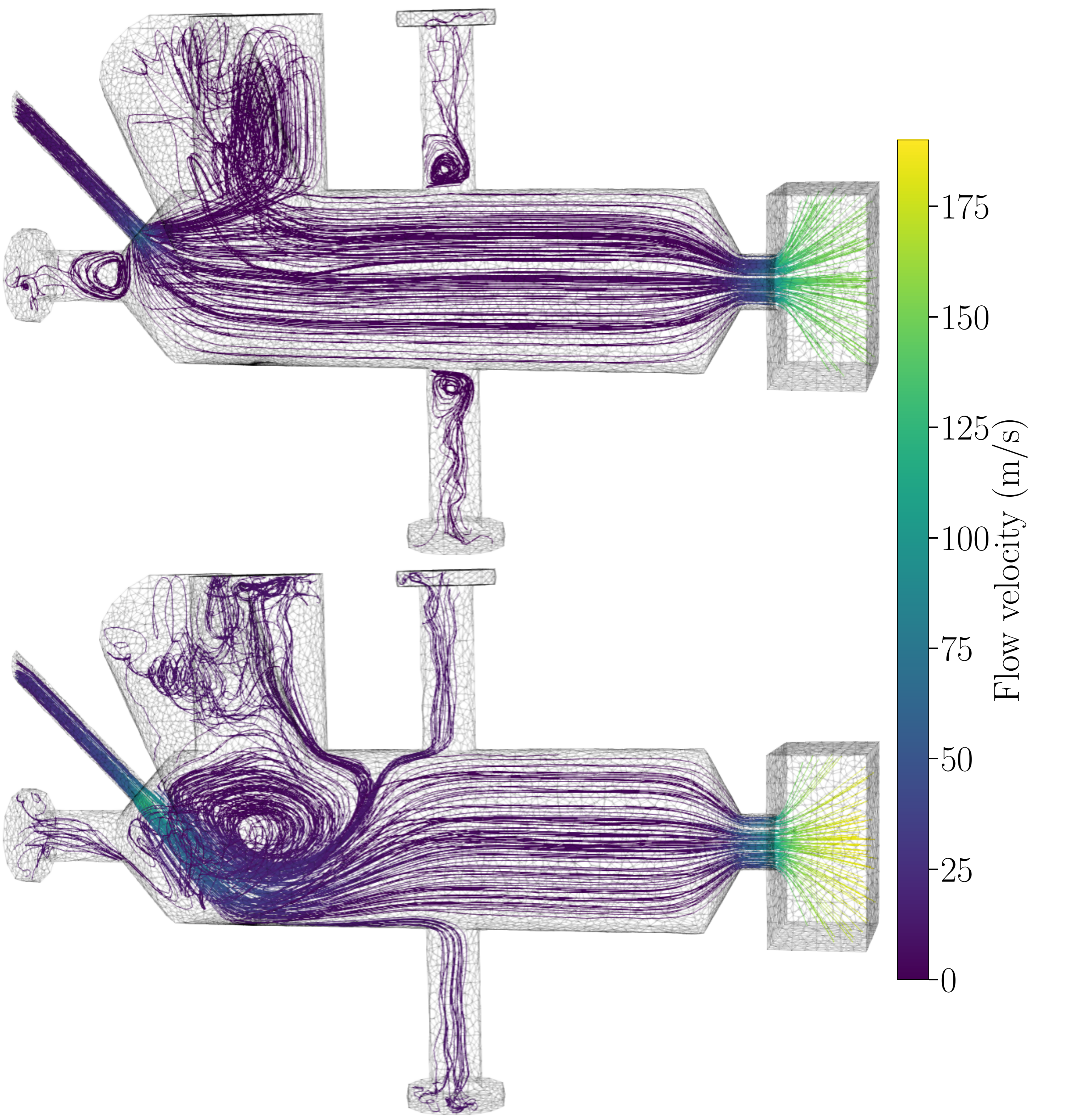}{(a)}{0}{(b)}{.5}
    \caption{
        Velocity field in the buffer-gas cell at flow rates of \SI{1}{SCCM} (a) and \SI{10}{SCCM} (b), visualized by individual particle trajectories. The trajectory lines indicate the local flow direction, with color representing the particles' velocities. For both flow rates, once in the cell main chamber, the particles flow toward the aperture, while recirculating patterns develop in the side chambers. In addition, for the large flow rate in (b), a large vortex forms near the helium inlet.}
    \label{fig:HeSim_vortex_formation}
\end{figure}

While the velocity field and density distributions characterize the macroscopic helium flow, the particle-based nature of the DSMC method also provides access to individual helium trajectories. Figure \ref{fig:HeSim_vortex_formation} shows such representative particle trajectories for two buffer-gas inflows. At higher inflows, starting around \SI{5}{SCCM}, a vortex develops in the main chamber directly at the helium inlet. This is consistent with observations in other numerical studies of buffer-gas cells~\cite{bulleidCharacterizationCryogenicBeam2013, vogeleyHydrodynamicEffectsCryogenic2025} and reflects the onset of more complex flow dynamics as the Knudsen number decreases. 

\subsection{Helium beam formation}
Having characterized the helium flow inside the cell, we now consider the extracted helium beam. Upon exiting the cell, the aperture acts as a nozzle, converting thermal energy into directed kinetic energy. As a result, the transverse temperature drops from \SI{4}{K} inside the cell to approximately \SI{3}{K} at \SI{2}{mm} downstream of the aperture, while the mean forward velocity increases.

\begin{figure}[t]
    \centering
    \labeledimage{0.44\textwidth}{(a)}{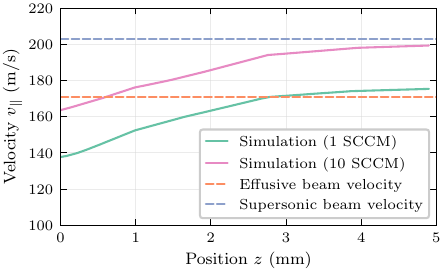}\\
    \labeledimage{0.44\textwidth}{(b)}{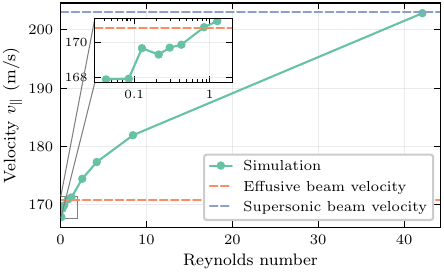}
    \caption{Beam formation dynamics. 
(a) Helium forward velocity $v_\parallel$ along the beam axis as a function of distance from the aperture exit, for buffer-gas flows of \SI{1}{SCCM} and \SI{10}{SCCM}. The helium atoms accelerate due to the buffer-gas density gradient behind the aperture. Dashed lines indicate the expected asymptotic velocities for the effusive and supersonic regimes~\cite{Hutzler2012}.
(b) Far-field forward velocity $v_\parallel$ as a function of Reynolds number $\mathrm{Re}$. The simulations transition from the effusive limit at $\mathrm{Re} \lesssim 1$ to the supersonic limit at large $\mathrm{Re}$, in agreement with the respective theoretical predictions~\cite{Hutzler2012}.
    }
    \label{fig:vel_boosting}
\end{figure}

This positional dependence of the mean forward velocity along the beam axis is further analyzed in Fig.~\ref{fig:vel_boosting}a. For our simulated flows, the helium beam approaches the fully ballistic regime at the edge of the simulation domain, as indicated by the flattening of the velocity curves. At \SI{1}{SCCM} the beam operates close to the effusive regime, with the asymptotic forward velocity approaching the effusive limit. At higher flows such as \SI{10}{SCCM} the beam is transitioning to the supersonic regime, resulting in a larger asymptotic forward velocity. We also observe that lower flows reach their asymptotic velocity closer to the aperture. 

The forward velocity $v_\parallel$ as a function of the Reynolds number $\mathrm{Re}$ is shown in Fig.~\ref{fig:vel_boosting}b. For $\mathrm{Re} \lesssim 1$ the forward velocity is roughly constant and consistent with the effusive prediction, while at larger $\mathrm{Re}$ it approaches the supersonic limit, in agreement with expectations~\cite{Hutzler2012}. 

Since the Knudsen number scales inversely with inflow, $\mathrm{Kn} \propto 1/f_\mathrm{in,b}$, the simulations span conditions from the free-molecular regime at low inflows to the continuum regime at high inflows.
Within a single simulation run, different regions of the cell can simultaneously be in different flow regimes --- near-free-molecular in the side chambers ($\mathrm{Kn} \gg 1$), transitional in the main chamber, and continuum-like at the inlet and aperture. 
Our results confirm that the PICLas framework correctly reproduces the expected helium flow behavior across all these regimes.


\section{Dynamics of the\\molecular trace species}
\label{sec:molecules}

Next, we introduce the molecular species into the simulation and examine the dynamics of plume expansion, thermalization, and beam formation. As mentioned above, the molecular species considered here is CaF.

In this section, simulations are performed within the background-gas approximation: The helium flow is taken from the final states of the simulations of Sec.~\ref{sec:helium}, and the molecules interact only with the buffer gas, but not with each other. Additionally, the helium flow in the simulation is unaffected by the molecules. The limitations of this approximation will be investigated in Sec.~\ref{sec:heating} below.

The molecules are initialized at the target boundary as described in Sec.~\ref{sec:framework}, modeling their creation by laser ablation. The ablation plume is represented as a thermal ensemble whose center-of-mass velocity is directed along the target surface normal~\cite{koolsGasFlowDynamics1992}. Particles are emitted according to a time-dependent surface flux following a $t^{-3}$ dependence over a duration of \SI{1}{ns}~\cite{taralloBaHMolecularSpectroscopy2016}, with an angular distribution proportional to $\cos^2\theta$, consistent with the cosine-power emission profiles commonly observed for laser-ablation plumes~\cite{koolsGasFlowDynamics1992,Konomi2009}. An initial translational temperature of \SI{10000}{K} is chosen, as observed in previous experimental studies of laser-ablated molecules~\cite{taralloBaHMolecularSpectroscopy2016}. The initial rotational and vibrational temperatures are set higher than the translational temperature, at \SI{12000}{K} and \SI{15000}{K}, respectively, consistent with the expectation that the internal degrees of freedom in ablation plasmas are typically excited to higher temperatures than the translational degrees of freedom~\cite{phillipsComparisonExcitationKinetic2023}.

Unless stated otherwise, the simulations presented in this section use a helium background with a density of approximately $n_{\mathrm{He}} \approx \SI{1.2e22}{m^{-3}}$, corresponding to the helium flow field obtained for a nominal inflow of \SI{10}{SCCM}. Roughly $5\times10^6$ simulation particles are inserted, each representing $10^8$ real molecules.

\subsection{Ablation plume expansion}

\begin{figure}[t]
    \centering
\includegraphics[width=0.43\textwidth]{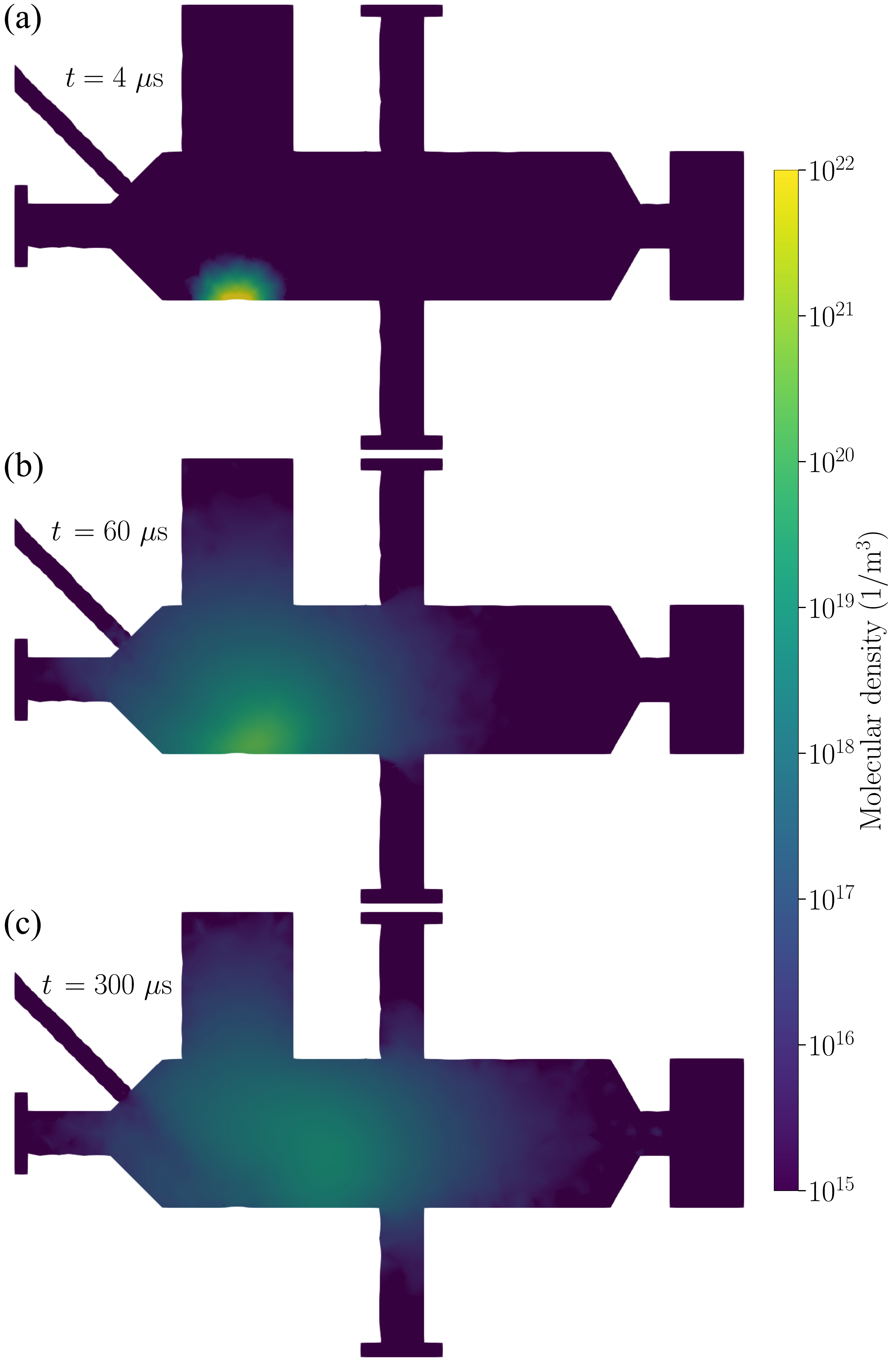}    
    \caption{
        Evolution of the molecular plume after the ablation pulse. We use a buffer-gas flow rate of \SI{10}{SCCM} and visualize the number density of the molecules at three sequential timesteps after the ablation. At early times (a), the plume expands along the target surface normal, driven by the velocities initialized in the ablation process. Over time (b,c), collisions with helium lead to diffusion and the bulk motion of the molecules is redirected toward the aperture by the buffer-gas flow. The decreasing total number of molecules reflects the losses to the cell walls via the absorbing boundary condition.
    }
    \label{fig:expansion_mol_plume}
\end{figure}

The time evolution of the molecular number density after their initialization is shown in Fig.~\ref{fig:expansion_mol_plume}. Immediately after injection, the molecular cloud expands away from the target along its surface normal, driven by the initial velocity distribution. Over time, collisions with helium randomize the molecular trajectories, and the helium flow begins to guide the plume toward the aperture. Most molecules eventually reach a cell wall and are lost via the absorbing boundary condition, with only a small fraction reaching the aperture. The fraction of molecules that are outcoupled depends sensitively on the initial velocity distribution: Molecules whose initial velocity vector points toward the aperture are more likely to exit the cell, while those directed toward the opposite wall are lost before the helium flow can redirect them. At lower buffer-gas densities, the plume propagates more rapidly toward the opposite wall, consistent with the expectation that helium flow plays a weaker role in redirecting the molecules at lower densities.

\subsection{Thermalization}

The molecular ensemble thermalizes with the helium buffer gas through elastic and inelastic collisions. Thermalization occurs for translational degrees of freedom as well as internal rovibrational degrees of freedom, all of which relax toward the \SI{4}{K} buffer-gas temperature through repeated collisions, although on very different timescales.

\begin{figure}[tbp!]
    \centering
    \labeledimage{0.43\textwidth}{(a)}{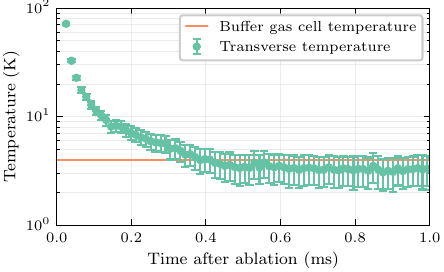}
    \hfill
    \labeledimage{0.412\textwidth}{(b)}{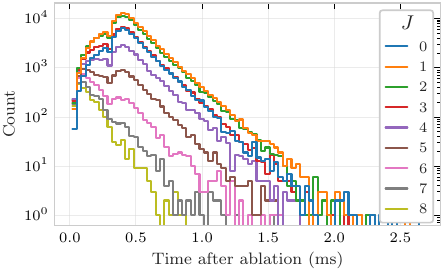}
    \hfill
    \labeledimage{0.412\textwidth}{(c)}{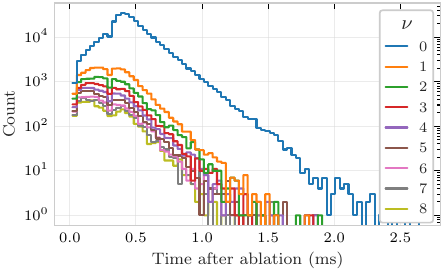}
    \caption{
    Thermalization of the molecular degrees of freedom.
    (a) Temperature $T$ inside the buffer-gas cell, extracted from Gaussian fits to the molecular velocity distributions in the transverse $x$- and $y$-directions. The motional temperature shows the characteristic early decay and asymptotic approach to the cell temperature (orange line), with translational thermalization complete after approximately \SI{0.4}{ms}.
    (b) Population of the first nine rotational levels $J$ as a function of time. Higher levels decay more rapidly, leaving only the lowest levels populated at later times. After roughly \SI{1}{ms} the ensemble has reached $T_\text{rot} \approx \SI{4}{K}$.
    (c) Population of the first nine vibrational levels $\nu$ as a function of time. Higher levels again decay first, though the vibrational thermalization proceeds more slowly than the rotational thermalization due to the lower per-collision relaxation probability. In all panels, the decreasing total particle count reflects losses to the cell walls via the absorbing boundary condition.
}
    \label{fig:thermalization}
\end{figure}

The translational thermalization is shown in Fig.~\ref{fig:thermalization}a. After a rapid decrease driven by the first few collisions, the motional temperature approaches the cell temperature asymptotically, as expected from kinetic theory~\cite{skoffDiffusionThermalizationOptical2011}.

As discussed previously, the relaxation of the internal degrees of freedom is modeled using fixed per-collision relaxation probabilities, which set how frequently an inelastic event occurs. When one does, the internal states are resampled via the Larsen-Borgnakke method~\cite{BORGNAKKE1975}, which guarantees convergence of the rovibrational distributions toward the local Boltzmann distribution, but does not implement state-specific selection rules. The following rotational and vibrational dynamics therefore illustrate the capabilities of this phenomenological relaxation model. They should not be interpreted as quantitative predictions of state-to-state population transfer.

The rotational relaxation probability of \SI{5}{\percent} is consistent with the order of magnitude of ab initio rotationally inelastic-to-elastic rate ratios calculated for CaF--He at cryogenic temperatures~\cite{londonoRotationalQuenchingMonofluorides2025}. 

The smaller vibrational relaxation probability of \SI{1}{\percent} reflects the slower vibrational relaxation expected in buffer-gas cooling. Unlike the rotational value, it is not calibrated against ab initio rates. Since vibrational quenching in cold helium is expected to be far less efficient than rotational quenching, the value used likely exceeds the true per-collision probability and should be regarded as an upper limit rather than an estimate.

The resulting evolution of the first nine rotational and vibrational levels is shown in Fig.~\ref{fig:thermalization}b,c. Higher levels decay first, as reflected in the steeper slopes of their time evolution, such that only the lowest levels remain populated at later times. After approximately \SI{1}{ms} the rotational ensemble has reached $T_\text{rot} \approx \SI{4}{K}$. The overall particle count decreases as a result of losses through the absorbing boundary condition on the cell walls and, to a lesser extent, through particles leaving the cell.

Within this relaxation model, the degree of vibrational cooling depends monotonically on the assumed per-collision probability. The value adopted here is therefore expected to overestimate the vibrational relaxation, and the simulated populations constitute an upper bound on the vibrational cooling achievable over the residence time in the cell: for any smaller probability, molecules leave the aperture with a vibrational distribution at least as hot. Even under this optimistic assumption, the extracted beam is still in the process of vibrationally relaxing when it exits the cell, long after the rotational degrees of freedom have fully thermalized to the cell temperature. 

\subsection{Extraction}
\label{sec:extraction}

\begin{figure}[tbp]
    \centering
    \labeledimagetwo{0.43\textwidth}{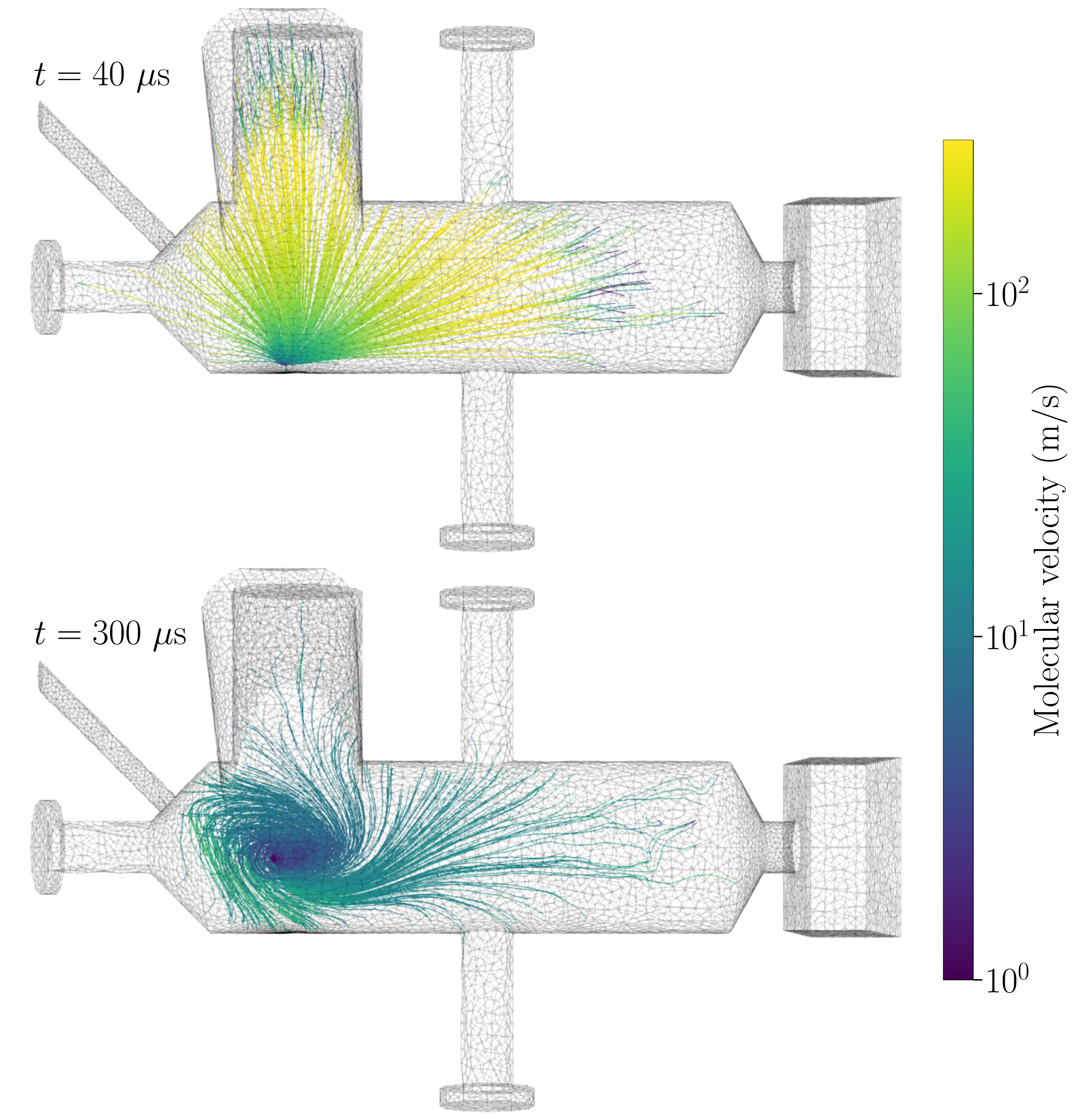}{(a)}{-0.01}{(b)}{0.49}
    \caption{
        Representative molecule trajectories at two stages of the simulation for a buffer-gas flow rate of \SI{10}{SCCM}. (a) Shortly after the ablation pulse, the molecule trajectories are dominated by the initial velocity conditions and most molecules propagate toward the wall opposite the target. (b) During thermalization, the molecular motion is increasingly guided by the helium flow. The molecules that reach the cell center subsequently undergo a random walk and most molecules are lost due to absorbing collisions with the cell walls before they reach the cell aperture.
    }
    \label{fig:CaF_trajectories}
\end{figure}

The extraction of molecules from the cell is a crucial process in buffer-gas cooling. Detailed insights into this process beyond bulk quantities can again be gained by following individual molecular trajectories.

Figure \ref{fig:CaF_trajectories} illustrates the evolution of the molecular trajectories from injection to extraction. At early times, most of the molecules propagate ballistically, effectively unaffected by the buffer gas. Over time, collisions increasingly randomize the trajectories until the molecules are coupled to the helium flow and begin to follow the buffer-gas flow trajectories. This continuous transition from ballistic to diffusive to flow-guided motion emerges naturally within the kinetic simulation framework. 

Notably, the mean forward velocity of the molecules in the cell center is $\bar{v}_z = \SI{47}{m\per s}$, which significantly exceeds the $\SI{5}{m\per s}$ helium bulk velocity inside the cell. This reflects the initialization of the molecules, which initially exhibit a directed velocity component from the ablation pulse and are only partially entrained in the helium flow at this stage. Collisions with the buffer gas continuously slow and redirect the molecules, but full coupling to the flow field is established only near the aperture, where the steep helium density gradient dominates the dynamics. 

The outcoupling efficiency --- the fraction of ablated molecules that exit the aperture --- is $
    \epsilon = N_\text{out}/N_\text{in} \approx \SI{0.014}{\percent},$
with only $N_\text{out} = 684$ of the $5\times10^6$ inserted simulation particles exiting the cell. We emphasize that $\epsilon$ is defined with respect to the total number of ablated molecules $N_\text{in}$, a quantity not directly accessible in experiments. In contrast, reported extraction efficiencies on the order of $10\%$ in experiments are typically referenced to the molecule number at a certain probe laser position~\cite{Truppe2023}. These ignore the dominant initial plume losses and are therefore not comparable. Sampling after this initial plume loss, we recover values closer to those in experiments, which are, however, inherently highly sensitive to the specific probing geometry. 

The observed low outcoupling efficiency reflects the dominance of wall losses. Most molecules travel directly toward the wall opposite the target before the helium flow can redirect them, and those that do thermalize subsequently undergo a random walk that also frequently leads to wall contact. 

Interestingly, angling the target by approximately $79^\circ$ such that its normal vector points toward the aperture more than doubles the outcoupling efficiency to \SI{0.033}{\percent}. This example demonstrates the sensitivity of the extraction efficiency to the source geometry and represents a first step toward the systematic optimization of cell designs using the simulation framework presented here.

\subsection{Beam formation}
As a final step, we turn to the properties of the beam itself, downstream of the output aperture of the cell.

When molecules pass through the aperture to form the molecular beam, their mean forward velocity increases to $\bar{v}_z = \SI{74}{m\per s}$ directly after the aperture and $\bar{v}_z = \SI{88}{m\per s}$ at \SI{2}{mm} downstream, driven by the longitudinal pressure gradient at the aperture. This increase in forward velocity downstream of the aperture is a signature of the hydrodynamic boosting characteristic of buffer-gas beam formation. As expected from the mass difference between CaF and helium, the molecular beam velocity remains well below the mean helium-beam velocity of $\bar{v}_{z,\text{He}} \approx \SI{145}{m\per s}$ at the same position. 

The simulated trajectories of the molecules that exit the cell and the corresponding velocity distributions further allow us to estimate the opening angle of the molecular beam outside the cell. For this, we sample the forward velocity distribution \SI{2}{mm} away from the aperture. The corresponding opening angle for the molecular beam is $\SI{35}{\degree} \pm \SI{4}{\degree}$.
This value is only slightly larger than the simplified prediction $\Delta \theta_\mathrm{CaF,He} \approx 2 \sqrt{m_\mathrm{He}/m_\mathrm{CaF}} \approx \ang{30}$ of Ref.~\cite{Hutzler2012}. The difference is consistent with the beam not being fully hydrodynamic and not yet fully ballistic.


\section{Beyond the background-gas approximation}
\label{sec:heating}

The results presented in Sec.~\ref{sec:molecules} were obtained within the usual background-gas approximation, in which the helium flow is fixed and the molecules are treated as a dilute perturbation that does not influence the buffer gas. This approximation simplifies the modeling considerably and is the basis of all existing hybrid simulation approaches. However, it is not inherent to the present kinetic framework and its validity is not guaranteed under certain experimental conditions involving laser ablation. 

\begin{figure}
    \centering
    \labeledimage{0.43\textwidth}{(a)}{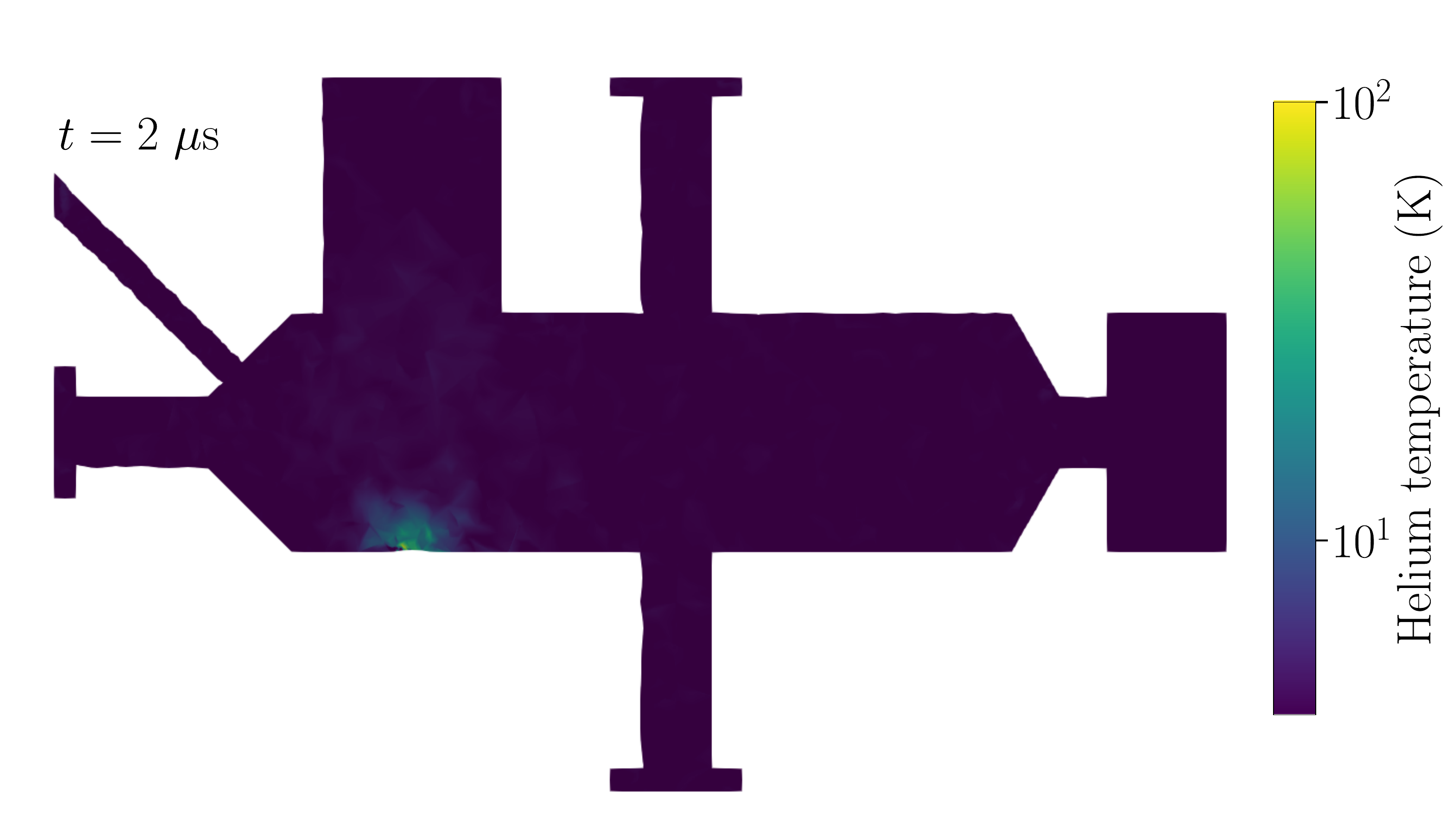}
    \hfill
    \labeledimage{0.43\textwidth}{(b)}{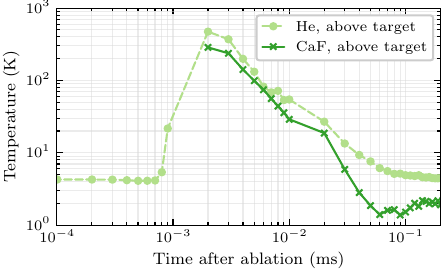}
    \caption{Energy transfer from a hot molecular plume to the buffer gas, for a full simulation beyond the background-gas approximation. The setup starts from the converged \SI{0.05}{SCCM} helium flow, inserting around $10^{14}$ real molecules at an initial transverse temperature of $T_\text{trans} = \SI{1000}{K}$. 
    (a) Local heating of the buffer gas induced by coupling to the molecular plume at \SI{2}{\mu s} after the laser ablation pulse. The heating is only present close to the ablation target.
    (b) Helium (circles, dashed line) and CaF (crosses, solid line) transverse temperatures \SI{0.2}{mm} above the target as a function of time after the ablation pulse. The helium temperature rises before the bulk of the molecular plume arrives and both species subsequently relax toward the cell temperature.}
    \label{fig:T_He_vs_t}
\end{figure}

To investigate this coupling, we perform a simulation without the background-gas approximation, starting from the converged \SI{0.05}{SCCM} helium inflow simulation. We insert approximately $10^{14}$ real molecules at an initial transverse temperature of $T_\text{trans}=\SI{1000}{K}$, several orders of magnitude above the \SI{4}{K} helium buffer gas. 

The reduced buffer-gas flow significantly reduces the computational cost of the fully coupled simulation while simultaneously making the molecular plume a stronger perturbation relative to the helium buffer gas than in the higher-flow simulations of Sec.~\ref{sec:molecules}. These parameters therefore allow the coupled helium--molecule dynamics to be isolated and analyzed clearly. 

As shown in Fig.~\ref{fig:T_He_vs_t}a, the interaction between the molecular plume and the helium buffer gas leads to significant but localized heating of the helium.

In Fig.~\ref{fig:T_He_vs_t}b we analyze these heating dynamics further and correlate the temperatures of molecules and helium as a function of time after ablation. The helium temperature increases before the bulk of the molecular plume arrives, which is a direct consequence of the high thermal velocity of the helium. The transverse temperature of the helium reaches approximately 470\,K, corresponding to a ten-fold increase in the mean thermal velocity of the buffer gas. The heated buffer-gas atoms redistribute energy throughout the cell faster than the heavier, slower molecules can propagate. Over time, both species cool back to the cell temperature in parallel, providing a self-consistent picture of the full thermalization process.

Since the plume is more perturbing here relative to the helium than in the higher-flow simulations of Sec.~\ref{sec:molecules}, the heating observed in this example is expected to overestimate the effect neglected by the background-gas approximation in those regimes. The approximation therefore remains well justified there, in particular since the perturbation caused by the heating stays local to the region around the ablation target.

The observed dynamics are reminiscent of the thermalization dynamics seen in experiments, where heating of the buffer gas by the ablation pulse has previously been inferred indirectly from the measured thermalization~\cite{Albrecht2020,skoffDiffusionThermalizationOptical2011}. Here, the fully kinetic simulation provides a microscopic picture of this process, revealing how energy is transferred from the molecular plume to the helium and subsequently redistributed through the buffer gas. However, many experimentally relevant conditions lead to substantially more complex coupled dynamics than those considered here, and quantifying the exact magnitude of the coupling there will require a systematic study of its dependence on ablation energy, buffer-gas density, and cell geometry, which is beyond the scope of the present proof-of-principle study.

The present example demonstrates that such coupled molecule--buffer-gas dynamics, inaccessible under the background-gas approximation, are captured self-consistently within the fully kinetic framework.


\section{Conclusion}

We have presented fully kinetic simulations of a cryogenic buffer-gas cell using the DSMC method implemented in the PICLas framework, treating the helium buffer gas and ablated molecules within a single unified model. The simulations reproduce characteristic features of buffer-gas sources across a wide range of operating
conditions, including the equilibration of the helium density, nozzle cooling at the aperture, vortex formation at high inflows, thermalization and directed transport of the molecular species, and beam extraction, divergence, and forward velocity. Rather than reproducing a single experimental realization, the present work benchmarks the simulation framework against the established phenomenology summarized in Ref.~\cite{Hutzler2012}, thus validating the underlying physical mechanisms over a broad range of experimentally observed buffer-gas beam behaviors. As a further proof of principle, we have demonstrated that the fully kinetic treatment can capture energy transfer from the hot ablation plume to the helium buffer gas --- an effect inaccessible to approaches operating under the background-gas approximation. These results establish DSMC-based kinetic simulations as a viable tool for systematic studies of cryogenic buffer-gas sources and provide a framework for exploring and optimizing cell geometries, buffer gases and molecular species.

\section*{Acknowledgements}
We thank Jesús Pérez-Ríos and Mateo Londoño for discussions and for sharing calculations of the CaF--He scattering cross-sections, and Marian Rockenh\"auser for contributions in the early stages of this work. The computational results have been achieved using the Austrian Scientific Computing (ASC) infrastructure. This research was funded in whole or in part by the Austrian Science Fund (FWF) 10.55776/PAT8306623. We acknowledge funding from the European Research Council (ERC) under Grant agreements No. 949431 (\mbox{NEWMAT}) and No. 899981 (MEDUSA). Views and opinions expressed are however those of the author(s) only and do not necessarily reflect those of the European Union or the European Research Council. Neither the European Union nor the granting authority can be held responsible for them. 

\bibliography{biblio}

\end{document}